\documentclass[aps,prb,preprint]{revtex4}%
\usepackage{amsfonts}
\usepackage{amsmath}
\usepackage{amssymb}
\usepackage{graphicx}%
\begin{document}
\title[Mechanical adiabatic gas relation]{A microscopic, mechanical derivation for the adiabatic gas relation}
\author{P. M. Bellan}
\affiliation{Applied Physics, California Institute of Technology, Pasadena CA 91125}

\begin{abstract}
It is shown that the ideal gas adiabatic relation, $PV^{\gamma}%
=\mbox{constant}$, can be derived by considering the motion of a particle
bouncing elastically between a stationary wall and a moving wall.

\end{abstract}
\maketitle



\section{Introduction}

The simplest form of the adiabatic gas relation is the observation that the
temperature of a thermally insulated gas increases when it is compressed and
decreases when it is expanded. According to the historical review of this
subject by Kuhn,\cite{Kuhn} the first publication documenting this behavior
was by the Scottish physician William Cullen in the mid 17th century.
Experimental observations were summarized by the relation $PV^{\gamma
}=\mbox{constant}$, where the exponent $\gamma$ was determined to exceed
unity. The deviation of $\gamma$ from unity is what allowed Sadi Carnot to
develop his eponymous cycle. (Apparently Carnot did not have the correct value
of $\gamma$, but the use of an incorrect value did not affect his fundamental
result that heat engine efficiency depends only on inlet and outlet temperatures.)

Serious attempts to develop a theoretical explanation for the adiabatic gas
relation were undertaken by Laplace, Poisson, and others in the early 19th
century, but no single individual has been identified as being the first to
provide the correct theoretical explanation. Since the mid-19th century
development of thermodynamics, the adiabatic gas relation has been established
from first principles using thermodynamic arguments. The standard
thermodynamic derivation is based on considering the temperature change of the
gas in a cylinder at constant pressure or constant volume while taking into
account specific heats at constant pressure and constant volume.\cite{serway}

The purpose of this paper is to show that the adiabatic gas relation
$PV^{\gamma}=\mbox{constant}$ is a direct consequence of an important property
of periodic mechanical motion, namely adiabatic invariance. Although the word
adiabatic is used in both the mechanical and thermodynamic contexts, its
meaning in the mechanical context differs from its meaning in the
thermodynamic context\cite{adiabatic} because the concept of heat does not
exist in the mechanical context. The derivation presented here provides
insight into the fundamental microscopic dynamics underlying adiabaticity. The
derivation will first be presented for molecules with no internal degrees of
freedom and then extended to molecules with internal degrees of freedom. Two
standard properties of an ideal gas will be invoked repeatedly, namely, the
properties of an ideal gas occupying a volume $V$ do not depend on the shape
of the volume, and collisions cause all properties of an ideal gas to become isotropic.

\section{Molecules without internal degrees of freedom}

A molecule with no internal degrees of freedom may be represented by a point
particle. Consider the situation sketched in Fig.~1 where a point particle
bounces back and forth with speed $v$ between two walls. The two walls are
denoted as 1 and 2 and are separated by a distance $L$. Wall 1 is stationary
while wall 2 approaches wall 1 with speed $u$ where $u<<v$. Both $u$ and $v$
are speeds, not velocities, that is, $u$ and $v$ are both positive quantities
and leftward motion is denoted by an explicit minus sign. Wall 2 therefore
moves in the lab frame with a velocity $-u$.%

\begin{figure}
[h]
\begin{center}
\includegraphics[
height=2.9954in,
width=5.2436in
]%
{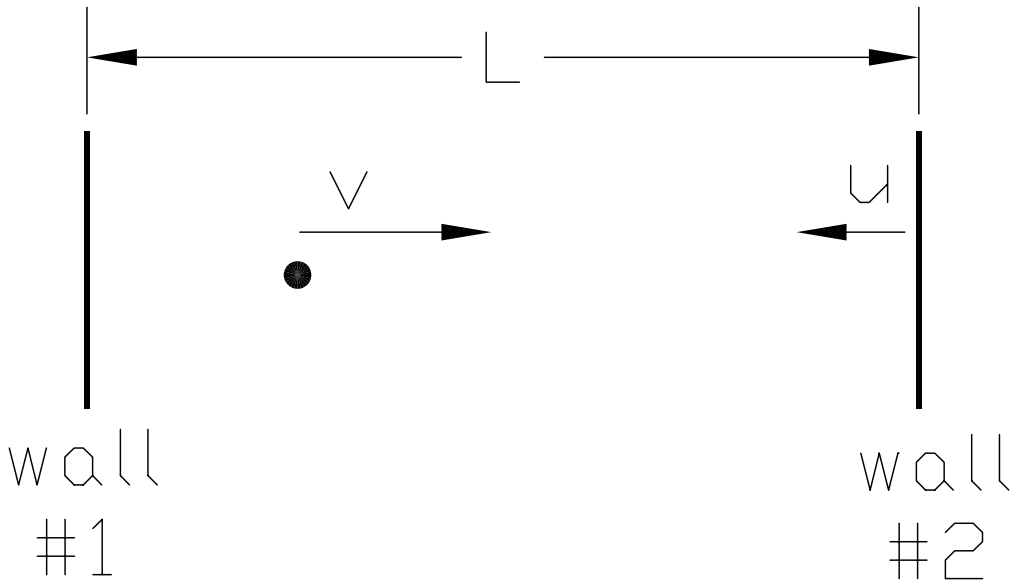}%
\caption{Wall 2 moves to the left with speed $u$. The particle has speed $v$.
The distance between the walls is $L$.}%
\end{center}
\end{figure}

The critical physical assumption is that the particle bounces elastically from
both walls. This microscopic assumption corresponds precisely to the
macroscopic adiabatic prescription that the gas is thermally insulated from
its surroundings. If particles bounced inelastically from the walls, the
temperature of the walls would change and the gas would not be thermally
insulated from the external world.

The transformation from the lab frame to the frame of wall 2 requires adding
$u$ to the lab frame velocity; the reverse transformation requires subtracting
$u$ from the velocity in the frame of wall 2. In the frame of wall 2, the
particle approaches wall 2 with velocity $v+u$ and after reflection has the
velocity $-v-u$. By transforming back to the lab frame, the lab frame velocity
after reflection is $-v-2u$. The change in the particle speed in the lab frame
due to reflection from wall 2 is therefore
\begin{equation}
\Delta v=2u. \label{deltaV}%
\end{equation}

Now consider the time for the particle to execute one cycle of bouncing
between walls 1 and 2. The distance traveled by the particle on starting from
wall 1, reflecting from wall 2, and then returning to wall 1 is $2L$, and so
the time to travel this distance is the bounce time
\begin{equation}
t_{b}=2L/v. \label{tb}%
\end{equation}
Equation~(\ref{tb}) can be used to calculate the change in inter-wall distance
during one complete bounce cycle. Because wall 2 is approaching wall 1, this
change is negative and is given by
\begin{equation}
\Delta L=-ut_{b}=-2Lu/v. \label{deltaL}%
\end{equation}
The presumption $u<<v$ combined with Eq.~(\ref{deltaL}) implies that $|\Delta
L|/L<<1$, and so we do not need to take into account changes in either $v$ or
in $L$ during a bounce cycle when calculating $t_{b}$; that is, $t_{b}$ is
well-defined even though both $L$ and $v$ change during a bounce cycle.

The combination of Eqs.~(\ref{deltaV}) and (\ref{deltaL}) gives the relation
\begin{equation}
L\Delta v+v\Delta L=0, \label{combine}%
\end{equation}
showing that $\Delta(vL)=0$, or
\begin{equation}
vL=\mbox{constant}. \label{adi}%
\end{equation}
The quantity $vL$ is called an adiabatic invariant\cite{Landau,Northrop} and
indicates conservation of the phase-space area enclosed by the phase-space
trajectory of periodic motion. This invariance is a fundamental property of
the periodic motion of Hamiltonian systems in which there exists a slowly
changing parameter (the word slowly in our case corresponds to $u<<v)$. Energy
is not conserved and scales as $v^{2}\sim1/L^{2}$.

Now assume that the particle bounces in three dimensions between the walls of
a cube with sides of length $L$, where, for each of the three dimensions, one
wall approaches its opposite partner as we have discussed. Thus, the cube
decreases in a self-similar fashion as shown in Fig.~2. Equation~(\ref{adi})
holds for each of the three dimensions and can be generalized as
\begin{equation}
v_{x}^{2}L^{2}=\mbox{constant},\ v_{y}^{2}L^{2}=\mbox{constant},\ v_{z}%
^{2}L^{2}=\mbox{constant}.\label{3Dcollisionless}%
\end{equation}
Hence,
\begin{equation}
v^{2}L^{2}=(v_{x}^{2}+v_{y}^{2}+v_{z}^{2})L^{2}%
=\mbox{constant}.\label{combine3Dcollisionless}%
\end{equation}
A large number of non-interacting particles in such a cube can be considered
as an ideal gas with temperature $T\sim$ ${<}v^{2}{>}$ where ${<}\ldots{>}$
denotes an average over all the particles. The average of
Eq.~(\ref{combine3Dcollisionless}) over all particles thus gives
\begin{equation}
TL^{2}=\mbox{constant}.\label{TL2}%
\end{equation}
Because the volume of the cube is $V=L^{3}$, Eq.~(\ref{TL2}) becomes
\begin{equation}
TV^{2/3}=\mbox{constant}.\label{adi2}%
\end{equation}
The ideal gas law gives the relation
\begin{equation}
PV=NkT,\label{ideal gas}%
\end{equation}
where $P$ is the pressure, $N$ is the number of particles, and $k$ is
Boltzmann's constant. If we use Eq.~(\ref{ideal gas}) to substitute for $T$ in
Eq.~(\ref{adi2}), we obtain the adiabatic relation
\begin{equation}
PV^{5/3}=\mbox{constant}.\label{pVgamma}%
\end{equation}
Equation~(\ref{pVgamma}) is the sought-after relation, but the derivation has
two important restrictions: the cube dimensions changed in a self-similar
fashion as in Fig.~2 and the particles were assumed to be non-interacting.%

\begin{figure}
[h]
\begin{center}
\includegraphics[
height=1.5641in,
width=3.9377in
]%
{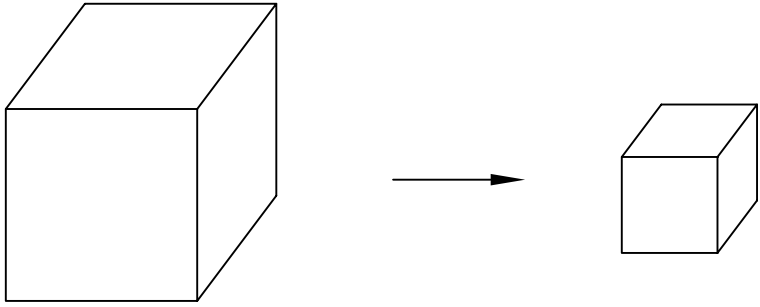}%
\caption{A cube with sides of length $L$ undergoing a self-similar decrease in
volume.}%
\end{center}
\end{figure}

We now repeat the derivation, but this time allow for collisions between
particles and also for non-self-similar changes in the dimensions of the
volume. As before, the volume is assumed to be thermally isolated from the
surroundings, which means that all reflections from the walls are
energy-conserving in the frames of the respective walls from which reflection
occurs. The combined assumptions of thermal isolation and existence of
interparticle collisions means that the final state after an arbitrary change
in the linear dimensions of the system is independent of how the change was
made. Thus, we are free to choose any sequence of steps we wish when going
from an initial state to some specified final state, provided that our two
assumptions stated above hold. The following discussion presents a sequence
chosen to elucidate the microscopic physics.

We start with a system that has an arbitrary initial shape, initial volume
$V_{0}$, initial temperature $T_{0}$; we want to determine the temperature $T$
when the volume changes to $V$. We first deform the shape of the volume
$V_{0}$ as shown in Fig.~3(a) so that the volume becomes rectangular with
linear dimensions $L_{x0}$, $L_{y0}$, and $L_{z0}$ with $V_{0}=L_{x0}%
L_{y0}L_{z0}$. This shape deformation at constant volume does not change the
initial gas temperature $T_{0}$ because ideal gas properties are independent
of the shape of the volume. Further, because the linear dimensions are
arbitrary, we set $L_{x0}$ sufficiently small that the bounce time in the $x$
direction is shorter than the nominal interparticle collision time $\tau_{c}$,
that is, we set $L_{x0}$ such that $2L_{x0}/v<<\tau_{c}$. Thus, particles
bounce many times in the $x$ direction before they make a collision with
another particle.%

\begin{figure}
[h]
\begin{center}
\includegraphics[
height=4.8152in,
width=2.6019in
]%
{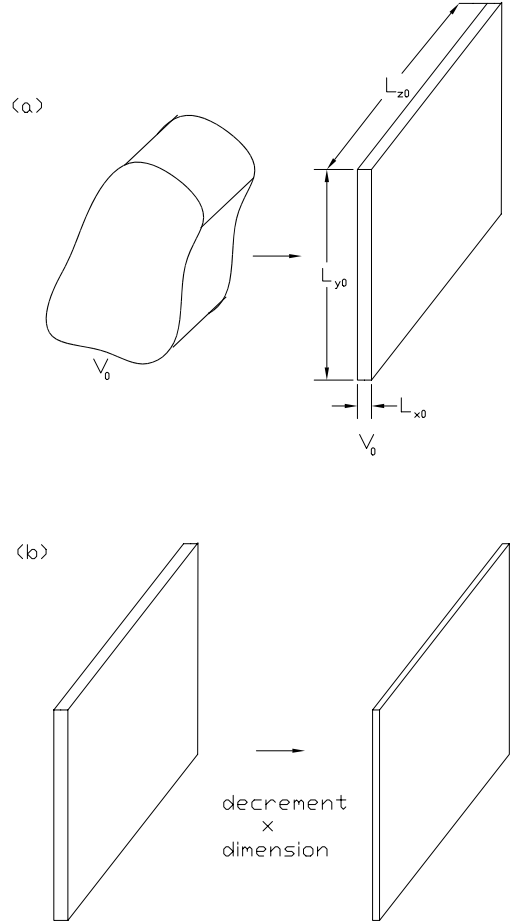}%
\caption{(a) Change of shape of the gas has no effect when the volume is kept
constant; the $x$-dimension is arranged to be so small so that the bounce time
in the $x$ direction is less than the collision time. (b) Decrease of
$x$-dimension only so that the instantaneous volume is proportional to $L_{x}%
$.}%
\end{center}
\end{figure}

Next we decrease just the $x$ dimension as shown in Fig.~3(b). Because
$dv_{x}=-v_{x}dL_{x}/L_{x}$, we may write
\begin{equation}
dv_{x}^{2}=-2v_{x}^{2}\frac{dL_{x}}{L_{x}}. \label{dvx2}%
\end{equation}
However, because only the $x$ dimension is being changed, $V\propto L_{x}$,
and so this collisionless change can be written as
\begin{subequations}
\label{13}%
\begin{align}
dv_{x}^{2} \big \vert _{\substack{\text{no} \\\text{collisions}}}  &
=-2v_{x}^{2}\frac{dV}{V}\label{split x}\\
dv_{y}^{2}\big\vert _{\substack{\text{no} \\\text{collisions}}}  &
=0\label{split y}\\
dv_{z}^{2}\big\vert _{\substack{\text{no} \\\text{collisions}}}  &  =0.
\label{energy split collisionless}%
\end{align}
Only the $x$-direction kinetic energy changes because the only dimension that
is being changed is $L_{x}$. Because there are no collisions, the $y$ and $z$
directions are decoupled from the $x$-direction, and so there are no changes
in the $y$- or $z$-direction kinetic energies.

After this decrease, we then wait a few collision times so that collisions
have a chance to equilibrate the change in kinetic energy among the three
directions. After collisions have shared the change in the $x$-direction
kinetic energy among all three dimensions, we obtain
\end{subequations}
\begin{subequations}
\begin{align}
dv_{x}^{2}\big\vert_{\substack{\text{after} \\\text{collisions}}}  &
=\frac{1}{3}dv_{x}^{2}\big\vert_{\substack{\text{no} \\\text{collisions}%
}}\label{split x after}\\
dv_{y}^{2}\big\vert_{\substack{\text{after} \\\text{collisions}}}  &
=\frac{1}{3}dv_{x}^{2}\big\vert_{\substack{\text{no} \\\text{collisions}%
}}\label{split y after}\\
dv_{z}^{2}\big\vert_{\substack{\text{after} \\\text{collisions}}}  &
=\frac{1}{3}dv_{x}^{2}\big\vert_{\substack{\text{no} \\\text{collisions}}}.
\label{energy split}%
\end{align}
The combined process of decreasing $x$ and then waiting for collisions to
equilibrate the energy gives, using Eqs.~(\ref{split x}) and
(\ref{split x after}),
\end{subequations}
\begin{equation}
dv_{x}^{2}=-\frac{2}{3}v_{x}^{2}\frac{dV}{V}, \label{dvx2
collisions}%
\end{equation}
or, upon integrating,
\begin{equation}
v_{x}^{2}V^{2/3}=\mbox{constant}. \label{vx2 collisions}%
\end{equation}
Because the collisions have equilibrated energy amongst the three dimensions
of motion, $v^{2}$ scales the same as $v_{x}^{2}$ and because $T$ scales as
$v^{2}$, we again obtain Eq.~(\ref{adi2}), which again leads to
Eq.~(\ref{pVgamma}). Once the system has attained its desired final volume, it
can be deformed at constant volume into whatever shape is desired. Thus, the
relation $PV^{5/3}=\mbox{constant}$ holds for any change in the volume and any
shape if there are collisions and if the system is thermally isolated.

\section{\textbf{Internal degrees of freedom}}

The analysis can now be extended to rigid diatomic and polyatomic molecules.
As is well-known, a rigid diatomic molecule has two additional degrees of
freedom due to rotational motion relative to the center of mass (there is
negligible rotational energy about the axis connecting the two atoms
comprising the molecule). On the other hand, a polyatomic molecule has three
additional rotational degrees of freedom because it can rotate in all three
directions about the center of mass. We let $n$ be the number of degrees of
freedom so $n=3$ for a monatomic molecule as discussed in Sec.~II, $n=5$ for a
diatomic molecule, and $n=6$ for a polyatomic molecule. The term
\textquotedblleft degrees of freedom\textquotedblright\ is defined as the
number of ways that collisions can share an increment in kinetic energy
supplied to one direction of motion amongst all possible types of motions of a
molecule. We define $T_{x}$, $T_{y},T_{z}$ as the kinetic energy due to center
of mass motion in the $x,y,z$ directions of a diatomic or polyatomic molecule,
that is,
\begin{equation}
T_{x}=\frac{M}{2}v_{x}^{2},\ T_{y}=\frac{M}{2}v_{y}^{2},\ T_{z}=\frac{M}%
{2}v_{z}^{2}, \label{Tx defn}%
\end{equation}
where $M$ is the molecular mass and $v_{x}$, $v_{y}$, and $v_{z}$ are center
of mass velocities. For a diatomic molecule, we define $I_{1}$ and $I_{2}$ as
the additional kinetic energies due to rotational motions relative to the
center of mass for the two allowed degrees of freedom of a diatomic molecule.
For a polyatomic molecule we similarly define $I_{1}$, $I_{2}$, $I_{3}$ as the
kinetic energies of the three allowed degrees of freedom due to rotational
motions relative to the center of mass. Consider now decreasing $L_{x}$ on a
time scale shorter than the collision time for a diatomic molecule. In this
case, Eqs.~(\ref{split x})-(\ref{energy split collisionless}) become
\begin{subequations}
\begin{align}
dT_{x}\big\vert_{\substack{\text{no}\\\text{collisions}}}  &  =-2T_{x}%
\frac{dV}{V}\\
dT_{y}\big\vert_{\substack{\text{no}\\\text{collisions}}}  &  =0\\
dT_{z}\big\vert_{\substack{\text{no}\\\text{collisions}}}  &  =0\\
\ dI_{1}\big\vert_{\substack{\text{no}\\\text{collisions}}}  &  =0\\
\ dI_{2}\big\vert_{\substack{\text{no}\\\text{collisions}}}  &  =0.
\label{diatomic no collisions}%
\end{align}

After the collisionless decrease, we wait for collisions to equilibrate the
change in kinetic energy amongst the five degrees of freedom and so obtain
\end{subequations}
\begin{subequations}
\begin{align}
dT_{x}\big\vert_{\substack{\text{after} \\\text{collisions}}}  &  =-\frac
{2}{5}T_{x}\frac{dV}{V}\\
dT_{y}\big\vert_{\substack{\text{after} \\\text{collisions}}}  &  =-\frac
{2}{5}T_{x}\frac{dV}{V}\\
dT_{z}\big\vert_{\substack{\text{after} \\\text{collisions}}}  &  =-\frac
{2}{5}T_{x}\frac{dV}{V}\\
dI_{1}\big\vert_{\substack{\text{after} \\\text{collisions}}}  &  =-\frac
{2}{5}T_{x}\frac{dV}{V}\\
dI_{2}\big\vert_{\substack{\text{after} \\\text{collisions}}}  &  =-\frac
{2}{5}T_{x}\frac{dV}{V}. \label{diatomic after collisions}%
\end{align}
If we repeat this process of decreasing collisionlessly, then waiting for
collisions to equilibrate the energy among the five degrees of freedom,
$T_{x}$ is seen to be governed by
\end{subequations}
\begin{equation}
dT_{x}=-\frac{2}{5}T_{x}\frac{dV}{V}, \label{dTx}%
\end{equation}
which may be integrated to give
\begin{equation}
T_{x}V^{2/5}=\mbox{constant}. \label{TxV52}%
\end{equation}
Because the temperature $T$ scales as $T_{x}$, we have
\begin{equation}
TV^{2/5}=\mbox{constant}. \label{TV5/2}%
\end{equation}

Clearly, Eq.~(\ref{TV5/2}) can be generalized to the form
\begin{equation}
TV^{2/n}=\mbox{constant}, \label{TV^n}%
\end{equation}
where $n=3$ for a monatomic molecule, $n=5$ for a diatomic molecule, and $n=6
$ for a polyatomic molecule. If we combine Eq.~(\ref{TV^n}) with
Eq.~(\ref{ideal gas}), we obtain the general result
\begin{equation}
PV^{\gamma}=\mbox{constant}, \label{genPVgamma}%
\end{equation}
where
\begin{equation}
\gamma=\frac{2+n}{n}. \label{gamma}%
\end{equation}
Thus $\gamma=5/3$ for a monatomic molecule, $\gamma=7/5$ for a diatomic
molecule, and $\gamma=4/3$ for a polyatomic molecule.

\section{Discussion}

Adiabatic gas behavior is closely related to the one-dimensional periodic
motion of a particle in a slowly changing system. This motion satisfies the
fundamental relation, $vL=\mbox{constant}$, which is a limiting form of the
general adiabatic conservation rule for slowly changing periodic motion of a
Hamiltonian system,\cite{Landau} namely that $\oint pdq=\mbox{constant}$,
where $p$ and $q$ are generalized coordinates and the integral is over one
period of oscillation. The $vL=\mbox{constant}$ relation is the basis for
Fermi acceleration,\cite{Fermi} the process by which cosmic charged particles
can be accelerated to extreme energies when they bounce back and forth between
converging magnetic fields playing the role of converging walls.

As mentioned, it is the \textit{elastic} bouncing of the particle from the
walls that makes the process adiabatic; if the bouncing were inelastic, the
wall would heat up in which case the gas would not be thermally isolated. The
$vL=\mbox{constant}$ relation also is intrinsic to the WKB approximation of a
simple harmonic oscillator, for example, a pendulum with slowly varying length
or equivalently the Schr\"{o}dinger equation for a particle in a slowing
changing potential well. When applied to the pendulum problem, the WKB
approximation involves solving the equation $d^{2}x/dt^{2}+\omega^{2}(t)x=0$,
where $\omega(t)$ is a time-dependent frequency.\cite{Griffiths} The frequency
$\omega(t)$ is essentially the inverse of the bounce time defined in
Eq.~(\ref{tb}) and so $\omega(t)\sim t_{b}^{-1}\sim v/L$. Because
$vL=\mbox{constant}$, it is seen that $v^{2}/\omega(t)\sim\mbox{constant}$ and
so the harmonic oscillator energy increases in proportion to the frequency.
Energy is not conserved, but the ratio of energy to frequency is.

When there are non-self-similar volumetric changes and collisions, our
derivation of the adiabatic gas law from mechanical considerations requires
separating the process into a sequence of infinitesimal steps where each step
has two successive sub-steps, namely a one-dimensional change in volume which
adiabatically (in the mechanical sense) changes the center of mass kinetic
energy of the changed dimension only, and then a collision-mediated sharing of
this energy change among all allowed degrees of freedom. This point of view
clearly shows why diatomic and polyatomic molecules have a different $\gamma$
from monatomic molecules.

Finally, it is worth addressing a possible confusion regarding whether
adiabatic processes are fast or slow: adiabatic processes are characterized
both ways, depending on the context. In order for an adiabatic invariant to
exist in the context of classical mechanics, the system must change slowly,
that is, the change of the bounce time per bounce must be small compared to
the bounce time.\cite{this} From the thermodynamic point of view, an adiabatic
process must be fast compared to the time for heat to leave the system, that
is, the change in the dimensions of the configuration must occur faster than
the time for walls to heat up due to inelasticities of the reflecting
particles. Possible confusion is avoided by realizing that adiabatic processes
must be slow compared to particle bounce times so as to satisfy the $u<<v$
assumption, but fast compared to heat loss times so as to satisfy the
assumption that the particle bounces elastically from the wall.

\begin{acknowledgments}
The author wishes to thank D.\ L.\ Goodstein for suggesting that non-monatomic
molecules ought to be considered. This work is supported in part by USDOE.
\end{acknowledgments}

\section*{}


\begin{thebibliography}{9}                                                                                                %


\bibitem {Kuhn}T. S. Kuhn, "The caloric theory of adiabatic compression," Isis
\textbf{49}, 132--140 (1958).

\bibitem {serway}See for example, R. A. Serway and R. J. Beichner,
\textit{Physics for Scientists and Engineers} (Brooks/Cole Publishing, 2000),
5th ed., Vol. 1, p. 650.

\bibitem {adiabatic}The word adiabatic means \textquotedblleft at constant
heat\textquotedblright\ in the context of thermodynamics and thus indicates
that the system is thermally insulated from the external world. Adiabatic does
not mean that energy is conserved within the system and adiabatic processes
typically involve energy changes of the system.

\bibitem {Landau}L. D. Landau and E. M. Lifshitz, \textit{Mechanics} (Pergamon
Press, Oxford, 1969), 2nd ed., p. 154.

\bibitem {Northrop}T. G. Northrop, \textit{The Adiabatic Motion of Charged
Particles} (Interscience, New York, 1963), p. 47.

\bibitem {Fermi}E. Fermi, "Galactic magnetic fields and the origin of cosmic
radiation," Astrophys. J. \textbf{119}, 1--6 (1954).

\bibitem {Griffiths}For a discussion of this in the context of the
Schr\"{o}dinger equation, see for example, D. J. Griffiths,
\textit{Introduction to Quantum Mechanics} (Prentice Hall, Upper Saddle River,
NJ, 1995), p. 274.

\bibitem {this}This slowness requirement is essentially the WKB criterion for
a variable-length pendulum. See, for example, Ref.~\onlinecite{Landau}.
\end{thebibliography}
\end{document}